\documentclass[12pt]{article}
\pdfoutput=1
\usepackage{jhep-mod}
\usepackage{bm}
\usepackage{amssymb,amsmath,amsthm}
\usepackage{mathrsfs} 
\DeclareMathOperator{\tr}{tr}
\def\I{{\mathbb{I}}}

\def\widebar{\overline}

\def\0{{\mathbf{0}}}

\def\mathsf{\tr}
\def\TODAY{Monday 18 June 2012}
\begin{document}
\title{Null Energy Condition violations in bimetric gravity}
\author{Valentina Baccetti, Prado Martin-Moruno, {\rm and} Matt Visser}
\affiliation{
School of Mathematics, Statistics, and Operations Research,\\
Victoria University of Wellington, PO Box 600, Wellington 6140, New Zealand}
\emailAdd{valentina.baccetti@msor.vuw.ac.nz}
\emailAdd{prado@msor.vuw.ac.nz}
\emailAdd{matt.visser@msor.vuw.ac.nz}
\abstract{
We consider the effective stress-energy tensors for the foreground and background sectors in ghost-free bimetric gravity.  By considering the symmetries
of the theory, we show that the foreground and background null energy conditions (NECs) are strongly anti-correlated. In particular, the NECs can only be simultaneously fulfilled when they saturate, corresponding to foreground and background cosmological constants. In all other situations, either the foreground or the background is subject to a NEC-violating contribution to the total stress-energy.


\bigskip
\noindent
\TODAY;  \LaTeX-ed \today.
}
\maketitle


\clearpage
\section{Introduction}
In 1970 Isham, Salam, and Strathdee hypothesized the existence of a spin-2 $f$-meson interacting with the spacetime metric $g$,  and with a kinetic 
term of the Einstein-Hilbert form~\cite{Isham:1971gm}.
This theory is variously known as bigravity, bimetric gravity, or $f$--$g$ gravity, and consists of two mutually interacting dynamical metrics.
The authors of this seminal paper noted that such a theory could have significant consequences in many different fields of theoretical physics,
leading to many interesting questions. For example, whether the gravity associated with this new $f$ metric could be repulsive for short distances
and, in this case, what would be the implications for black hole physics~\cite{Isham:1971gm}.

The problem with bimetric gravity was that it is generally affected by the same ghost instability appearing in massive 
gravity~\cite{Boulware:1973my}, a circumstance which had severely constrained interest in the model. 
However, a ghost-free bimetric gravity theory has recently been presented by Hassan and Rosen in reference~\cite{Hassan:2011zd}.
The construction of such a theory has been possible due to a quickly developing research programme. 
This programme started by first showing that there is a massive gravity theory which is ghost-free in the decoupling 
limit \cite{deRham:2010ik}, and even up to fourth order in non-linearities \cite{deRham:2010kj}. 
Second,  the theory considered in references~\cite{deRham:2010ik, deRham:2010kj} was generalized to allow general background metrics~\cite{Hassan:2011vm}. 
Later on, it was shown that this massive gravity theory in a general background is in fact ghost-free beyond the decoupling limit~\cite{Hassan:2011hr, Hassan:2011tf, Hassan:2011ea, Hassan:2012qv}. (See also~\cite{Kluson:2012wf} and references therein.)
The consistency of the theory is maintained when one gives dynamics to the background metric $f$~\cite{Hassan:2011zd}, although the underlying philosophy of the theory is completely changed~\cite{Baccetti:2012bk}.

In this context, collectively we find ourselves retracing the route of Isham, Salam, and Strathdee, in some sense coming back to propose 
the  same type of questions they considered. 
In fact, from a cosmological point of view, the scientific community has now gone further, not only trying to understand possible repulsive 
effects related to the new metric, but even asking whether such effects could affect large-distance physics in  our own gravitational 
sector.
Thus, bimetric gravity cosmologies~\cite{vonStrauss:2011mq, Volkov:2011an, Comelli:2011zm, Baccetti:2012bk}
have been considered in an attempt to explain the apparent accelerated expansion of our universe in the current epoch.
On the other hand, effects in black hole physics have also been studied~\cite{Comelli:2011wq, Banados:2011hk, Volkov:2012wp}, 
though it should be noted that some conclusions can also be extracted from general considerations and basic 
symmetry assumptions~\cite{Deffayet:2011rh}.

We shall demonstrate that there is a general way to in some sense classify the nature of the foreground-background gravitational 
interaction that we are facing.
As is well known, the classical energy conditions~\cite{Hochberg:1998ha, Hochberg:1998qw, Barcelo:2002bv, w-matter} establish the gravitational properties that one would 
expect to be fulfilled
by common classical materials. Thus, one could usefully propose that one way to understand the gravitational properties of a given 
theory which
modifies general relativity, is to consider whether the effects associated to the modifications might be equivalent to the presence
of some matter content fulfilling (or violating) the classical energy conditions. In the particular context of bimetric gravity 
one could pose the question: 
Does the fulfillment or violation of the null energy condition (NEC), the least restrictive 
(and hence the most powerful) of the classical energy conditions, in one gravitational sector imply fulfillment or violation of this condition in the other sector?
In this paper, we will give a full answer to this question.

The paper is organized as follows: In section~\ref{sT} we briefly summarize some previous results, fixing the
notation used through the paper. In section~\ref{stensors} we include some formal considerations about the symmetries of the theory
and their consequences.
We show that the null energy conditions of both spaces are strongly anti-correlated in section~\ref{sNEC}.
In section~\ref{discussion} we discuss our results. Some purely mathematical computations are then relegated to appendices \ref{squareroots}
and \ref{theorem}, whereas we generalize our results to a $n$-dimensional theory in appendix \ref{ndimension}.

\section{Bimetric gravity}\label{sT}
The action of bimetric gravity can be expressed quite generally as~\cite{Baccetti:2012bk}:
\begin{eqnarray}\label{actionbg}
S&=&-\frac{1}{16\pi G}\int d^4x\sqrt{-g} \left\{R(g) + 2\,\Lambda-2\,m^2 L_\mathrm{int}(g, f)\right\} +S_{({\rm m})}
\nonumber\\
&&
-\frac{\kappa}{16\pi G}\int d^4x\sqrt{-f} \left\{\widebar{R} (f)  +2\, \widebar\Lambda \right\} +\epsilon\,\widebar S_{({\rm m})},
\end{eqnarray}
with $S_{({\rm m})}$ and $\widebar S_{({\rm m})}$ the usual matter actions, with foreground and background matter fields coupled only to 
the foreground and background metrics $g_{\mu\nu}$ and $f_{\mu\nu}$, respectively. All interactions between these two sectors are confined 
to the term $ L_\mathrm{int}(g, f)$, which is an algebraic function of $g$ and $f$.\footnote{ 
We can recover the action of massive gravity by considering $\kappa=\epsilon=0$~\cite{Baccetti:2012bk}. In this case,
we would have an aether theory in which the dynamics of the physical metric $g_{\mu\nu}$ depends on a now non-dynamical
background metric $f_{\mu\nu}$.} The action (\ref{actionbg}) is ghost-free if the interaction term can be written as a linear
combination of the elementary symmetric polynomials of the eigenvalues of the  matrix $\gamma$ \cite{Hassan:2011zd},
where this matrix and the associated polynomials are defined through
\begin{equation}\label{gamma}
 \gamma^{\mu}{}_{\sigma}\gamma^{\sigma}{}_{\nu}=g^{\mu\sigma}f_{\sigma\nu},
 \qquad \hbox{that is} \qquad 
 \gamma^\mu{}_\nu = \left\{ \sqrt{g^{-1} f} \right\}^\mu{}_\nu,
\end{equation}
and 
\begin{equation}\label{defe}
\sum_{i=0}^4 \lambda^i \; e_i(\gamma) = \det(\I+\lambda \gamma).
\end{equation}
We can then, in 3+1 dimensions, express the interaction Lagrangian as~\cite{Baccetti:2012bk}
\begin{equation}\label{Lint}
 L_\mathrm{int}=\alpha_1\,e_1(\gamma)+\alpha_2\,e_2(\gamma)+\alpha_3\,e_3(\gamma),
\end{equation}
where
\begin{eqnarray}\label{symm}
 e_1(\gamma) &=&\tr[\gamma];\\
e_2(\gamma) &=&\frac{1}{2}\left(\tr[\gamma]^2-\tr[\gamma^2]\right);\\
e_3(\gamma) &=&\frac{1}{6}\left(\tr[\gamma]^3-3\tr[\gamma]\tr[\gamma^2]+2\tr[\gamma^3]\right).
\end{eqnarray}
The two remaining non-vanishing polynomials, $e_0(\gamma)=1$ and $e_4(\gamma)=\det(\gamma)$, have been absorbed into the kinetic terms of
$g_{\mu\nu}$ and $f_{\mu\nu}$, respectively --- because they lead to an effect on the equations of motion which is equivalent to a 
cosmological constant associated with each respective metric.
 If one additionally requires that
the coefficient of the mass term, which would appear multiplying $e_2(\I-\gamma)$, should be of the canonical Fierz--Pauli form,
then, using the expressions of reference~\cite{Hassan:2011vm}, which relate the coefficients
appearing in the interaction term written as a function of $\I-\gamma$ to the coefficients of $L_\mathrm{int}$ as expressed
in terms of $\gamma$; or, equivalently, using the
shifting theorem of reference~\cite{Baccetti:2012bk}, one has
\begin{equation}\label{relation}
 \alpha_1+2\alpha_2+\alpha_3=-1.
\end{equation}
It must be emphasized that in this theory both metrics have exactly the same status. Although the interaction term could naively seem to favor one of the
metrics over the other,  this is not really the case, as it fulfills the reciprocity relation~\cite{Hassan:2011zd, Baccetti:2012bk}
\begin{equation}\label{intbg}
\sqrt{-g} \; L_\mathrm{int}(\gamma) = \sqrt{-g} \; \sum_{i=0}^4 \alpha_i \; e_i(\gamma) =  \sqrt{-f} \; \sum_{i=0}^4 \alpha_{4-i} \; e_i(\gamma^{-1})
=\sqrt{-f} \; \widebar L_\mathrm{int}(\gamma^{-1}) .
\end{equation}
That is, the entire theory could be equivalently re-expressed using $f$ as the foreground metric and $g$ as the background.
We will use the terminology $f$-space and $g$-space throughout the paper to emphasize this equivalence.

By varying the action (\ref{actionbg}) with respect the two metrics, we obtain two sets of equations of motion. 
These are~\cite{Baccetti:2012bk}
\begin{equation}\label{motiong}
 G^{\mu}{}_{\nu}-\Lambda \,\delta^\mu{}_\nu =m^2\,T^{\mu}{}_{\nu}+8\pi G \;T^{({\rm m})\mu}{}_{\nu},
\end{equation}
and
\begin{equation}\label{motionf}
 \kappa\,\left(\widebar{G}^{\mu}{}_{\nu}-\widebar\Lambda\,\delta^\mu{}_\nu \right) =
m^2\, \widebar{T}^{\mu}{}_{\nu} +\epsilon\,8\pi G \,\widebar T^{({\rm m})\mu}{}_{\nu} ,
\end{equation}
where
\begin{equation}\label{Tg}
 T^{\mu}{}_{\nu}=\tau^{\mu}{}_{\nu}-\delta^{\mu}{}_{\nu}\,L_\mathrm{int},
\end{equation}
with 
\begin{equation}\label{taudef}
 \tau^{\mu}{}_{\nu}=\gamma^\mu{}_\rho\,\frac{\partial L_\mathrm{int}}{\partial \gamma^{\nu}{}_{\rho}},
\end{equation}
and
\begin{equation}\label{Tf}
 \widebar{T}^{\mu}{}_{\nu}=-\frac{\sqrt{-g}}{\sqrt{-f}}\;\tau^{\mu}{}_{\nu}.
\end{equation}
The indices of equation (\ref{motiong}) and (\ref{motionf}) must be raised and lowered using $g$ and $f$, respectively. 
Thus, the equations of motion of the $g$-space ($f$-space) are modified with respect to those
of general relativity by the introduction of an \emph{effective stress-energy } tensor associated to the interaction between 
the two geometries (and the quantity $\kappa/\epsilon$). 
Due to the invariance under diffeomorphisms of both matter actions in (\ref{actionbg}),
both effective stress-energy  tensors should fulfill the Bianchi-inspired constraints
\begin{equation}\label{Bianchi}
\nabla_\mu T^{\mu}{}_{\nu}=0; \qquad \widebar{\nabla}_\mu \widebar{T}^{\mu}{}_{\nu}=0.
\end{equation}
As has been pointed out in~\cite{vonStrauss:2011mq, Volkov:2011an}, once one constraint is enforced, for example $\nabla_\mu T^{\mu}{}_{\nu}=0$, then the other
is also automatically fulfilled.

Note that the modifications to the two equations of motion are very closely related. In fact everything can be expressed in terms of  
a single mixed-index tensor $ \tau^{\mu}{}_{\nu}$. 
Taking into account equations (\ref{Lint})
and (\ref{taudef}), 
and the explicit expressions for the derivatives of $e_1(\gamma)$, $e_2(\gamma)$,  and $e_3(\gamma)$, 
(either obtained by brute force or as deduced in appendix~\ref{theorem}),
 we see that $\tau^{\mu}{}_{\nu}$ can be written as a polynomial in the matrix $\gamma$. Specifically
\begin{eqnarray}\label{stress}
 \tau^{\mu}{}_{\nu}=
 \left(\alpha_1+\alpha_2\,e_1(\gamma)+\alpha_3\,e_2(\gamma)\right) \gamma^\mu{}_\nu -
\left(\alpha_2+\alpha_3\,e_1(\gamma)\right) \{\gamma^2\}^\mu{}_\nu +
\alpha_3 \{\gamma^3\}^\mu{}_\nu .
 \end{eqnarray}

\section{The two gravitational sectors}\label{stensors}

As already emphasized, in bimetric gravity the choice of $f$ as the background and $g$ as foreground metric, or vice versa,
is a matter of taste, since the action can be written equivalently (\ref{intbg}).
Such a symmetry between both gravitational sectors must still be present when one considers physical quantities as
the effective stress-energy tensor. In fact, one can make the symmetry explicit by considering equations (\ref{Tg}) and (\ref{Tf}) and writing
\begin{equation}\label{tensors}
\sqrt{-g} \; T^{\mu}{}_{\nu} + \sqrt{-f} \; \widebar T^{\mu}{}_{\nu} = - \sqrt{-g} \; L_\mathrm{int}(\gamma) \;\delta^{\mu}{}_{\nu}  = -\sqrt{-f} \; \widebar L_\mathrm{int}(\gamma^{-1})\;  \delta^{\mu}{}_{\nu} .
\end{equation}

On the other hand, the dependence of the interaction Lagrangian on the two metrics only through the square root matrix $\gamma$
ensures the fulfillment of other symmetries. In particular, the properties of matrix square roots (see appendix \ref{squareroots} for details)
allow us to note that the quantities
\begin{equation}\label{simetria1}
g_{\mu\rho}\{(\sqrt{g^{-1}f})^n\}^\rho{}_\nu; \qquad\hbox{and} \qquad f_{\mu\rho}\{(\sqrt{g^{-1}f})^n\}^\rho_\nu,
\end{equation}
are necessarily symmetric. In fact, this property is precisely what is guaranteeing that the effective stress-energy
tensors are symmetric. That can be noted lowering the indices of (\ref{Tg}) and (\ref{Tf}),
\begin{equation}\label{Tgabajo}
 T_{\mu\nu}=g_{\mu\sigma}\tau^{\sigma}{}_{\nu}-g_{\mu\nu}\,L_\mathrm{int},
\end{equation}
\begin{equation}\label{Tfabajo}
 \widebar{T}_{\mu\nu}=-\frac{\sqrt{-g}}{\sqrt{-f}}\;f_{\mu\sigma}\tau^{\sigma}{}_{\nu},
\end{equation}
and taking into account that equation~(\ref{stress}) implies $T_{\mu\nu}\propto \sum g_{\mu\rho}\{(\sqrt{g^{-1}f})^n\}^\rho{}_\nu$
and $ \widebar{T}_{\mu\nu}\propto \sum f_{\mu\rho}\{(\sqrt{g^{-1}f})^n\}^\rho_\nu$.

Moreover, we can also easily relate the null vectors of $f$ to those of $g$.
Let us consider a null vector with respect to $g$, so $g_{\mu\nu} \, k^\mu k^\nu = 0$.
One can then define 
\begin{equation}\label{defnull}
 \widebar k^\mu = \{\gamma^{-1}\}^\mu{}_\sigma k^\sigma.
\end{equation}
Contracting the indices of $f$ with $\widebar k^\mu$,
\begin{equation}\label{nulos}
f_{\mu\nu} \; \bar k^\mu \, \bar k^\nu =  k^\sigma \; \{\gamma^{-1}\}^\mu{}_\sigma f_{\mu\nu}  \{\gamma^{-1}\}^\nu{}_\rho \; k^\rho.
\end{equation}
But, the symmetry of the quantities (\ref{simetria1}) implies
\begin{equation}\label{simetria2}
f_{\mu\nu} = \gamma^\sigma{}_\mu \; g_{\sigma\rho} \; \gamma^\rho{}_\nu,
\end{equation}
which can be inverted, and inserted in (\ref{nulos}), leading to
\begin{equation}
f_{\mu\nu} \bar k^\mu \bar k^\nu = g_{\mu\nu} k^\mu  k^\nu = 0.
\end{equation}
Therefore, $\widebar k^\mu$ is a null vector with respect to $f$. 
Thus, by using $\gamma$ there is a 1-to-1 mapping between null vectors of $f$ and those of $g$.
As we show in the next section, this relation is extremely powerful.

\section{Null Energy Condition}\label{sNEC}
Let us now consider whether the modification of general relativity due to bimetric gravity would lead to effective stress-energy  tensors,
 $T^{\mu}{}_{\nu}$ or $\widebar{T}^{\mu}{}_{\nu}$, with characteristics similar to those describing classical common forms 
 of matter --- or might they violate the null energy condition instead?
In the first place, we will consider the effects in $g$-space. In this space, the NEC is the statement
\begin{equation}\label{NEC}
 T_{\mu\nu} \; k^\mu k^\nu\geq0,
\end{equation}
where $k^\mu$ is a null vector with respect to $g_{\mu\nu}$. The cosmological constant contribution to this stress
energy tensor (\ref{Tg})  will, of course, not affect the results regarding the NEC. Thus, we can express the NEC as
\begin{equation}\label{NECgtau}
 k^\mu\, (g_{\mu\sigma}\,\tau^{\sigma}{}_{\nu})\,k^\nu\geq0; \qquad\hbox{that is} \qquad  k_\mu (\tau^{\mu}{}_{\nu})\,k^\nu\geq0. 
\end{equation}
Taking into account equation (\ref{stress}) the NEC would be fulfilled in  $g$-space if
\begin{equation}\label{NECg1}
  \left(\alpha_1+\alpha_2\,e_1(\gamma)+\alpha_3\,e_2(\gamma)\right) [k_\mu\gamma^\mu{}_\nu k^\nu]-
\left(\alpha_2+\alpha_3\,e_1(\gamma)\right) [k_\mu\{\gamma^2\}^\mu{}_\nu k^\nu]+\alpha_3 [k_\mu\{\gamma^3\}^\mu{}_\nu k^\nu] \geq0.
\end{equation}
It must be emphasized that we are applying this condition to an effective stress-energy  tensor and not to any physical
source of matter. This effect is produced by the presence of the interaction with a second dynamical space.

In the second place, we can also study whether the NEC would be fulfilled
in $f$-space. Thus, considering a null vector $\widebar{k}$, now null with respect to $f_{\mu\nu}$,  the NEC in $f$-space can be written as
\begin{equation}\label{NECf}
  \widebar{T}_{\mu\nu}\; \widebar{k}^\mu\widebar{k}^\nu\geq0.
\end{equation}
Taking into account equation (\ref{Tf}), and noting that we are interested only in the \emph{sign} of this quantity (and not in its value),
this inequality leads to
\begin{equation}\label{NECftau}
\widebar{k}^\mu \,(f_{\mu\sigma}\,\tau^{\sigma}{}_{\nu})\,\widebar{k}^\nu\leq0.
\end{equation}
Replacing Equation (\ref{stress}), this is
\begin{equation}\label{NECf1}
  \left(\alpha_1+\alpha_2\,e_1(\gamma)+\alpha_3\,e_2(\gamma)\right) [\widebar{k}_\mu\gamma^\mu{}_\nu\widebar{k}^\nu]-
\left(\alpha_2+\alpha_3\,e_1(\gamma)\right) [\widebar{k}_\mu\{\gamma^2\}^\mu{}_\nu\widebar{k}^\nu]+\alpha_3 [\widebar{k}_\mu\{\gamma^3\}^\mu{}_\nu\widebar{k}^\nu]\leq0,
\end{equation}

On the other hand, as we have proven in the previous section, given a null vector with respect to
the metric $g$, $k^\mu$, we can always write a null vector with respect to $f$, $\widebar k^\mu$ by equation~(\ref{defnull}).
Therefore, we can write the terms appearing in
equation~(\ref{NECf1}) using expression~(\ref{defnull}) as
\begin{equation}
\widebar k_\mu \{\gamma^n\}^\mu{}_\nu \widebar k^\nu 
= k^\alpha \{\gamma^{-1}\}^\mu{}_\alpha  f_{\mu \sigma}  \{\gamma^{n-1}\}^\sigma{}_\nu k^\nu,
\end{equation}
which, taking into account equation~(\ref{simetria2}), leads to
\begin{equation}\label{violation}
\bar k_\mu \{\gamma^n\}^\mu{}_\nu \bar k^\nu = k_\mu \{\gamma^n\}^\mu{}_\nu k^\nu. 
\end{equation}
In view of expressions~(\ref{NECg1}) and (\ref{NECf1}), equation~(\ref{violation}) implies
that for every $k$ such that (\ref{NECg1}) is strictly satisfied, there is a $\widebar k$ such that (\ref{NECf1}) is violated, and 
vice versa. (The exceptional case is where both NECs are saturated.) 

It can be noted, from equations (\ref{NECgtau}), that the NEC in the $g$-space only
saturates if $\tau^\mu{}_\nu \propto \delta^\mu{}_\nu$, which is equivalent to the contribution of a cosmological constant;
but, through equation (\ref{NECftau}), this implies that the NEC in the
$f$-space also saturates. Therefore, the fulfillment of both NECs is only
possible if the contribution of the effective stress-energy tensors is equivalent to that of foreground and background cosmological constants.

\section{Discussion}\label{discussion}

We have considered bimetric gravity, that is, the theory which modifies general relativity by introducing a second
dynamical metric with the same status as that which governs the observable gravitational phenomena of ``our universe'', and  have studied the nature of the gravitational effects due to the existence of this second metric.

In the first place, we have gone into the implications of the existence of this equally preferred metric on the form of the 
effective stress-energy tensors which can be defined by gathering together the new terms appearing in the equations of motion.
Moreover, we have shown how the null vectors in both spaces can be easily related.

Nevertheless, the principal result of this paper refers to the NEC.
As in this theory there are two spacetime geometries, $g$-space and $f$-space, the NEC associated with the respective effective 
stress-energy  tensors can be studied in both spaces. We have shown that the expressions can be greatly simplified until arriving to
one surprising conclusion: {\it both NECs can be simultaneously satisfied if and only if the effect of the modification of general relativity
is equivalent to  foreground and background cosmological constants}. In any other situation the NEC is violated in one space or the other. Even more, as shown in appendix
\ref{ndimension}, this conclusion can be obtained independently of the dimension of spacetime.

We want to emphasize that we are considering the NEC associated to an effective stress-energy  tensor, not to real physical
matter. The understanding of the violation of the NEC associated to real physical matter or to modified theories of gravity is very different,
although they can lead to the occurrence of similar phenomena. In a cosmological context, it could lead to some kind of phantom cosmologies,
opening the door to the associated possible doomsdays, as the big rip \cite{Caldwell:2003vq} or big freeze 
\cite{BouhmadiLopez:2006fu,BouhmadiLopez:2007qb} future singularities \cite{Cattoen:2005dx}.
In an astrophysical framework, it would potentially allow the existence of wormholes in one of the gravitational sectors~\cite{Hochberg:1998ha, Hochberg:1998qw}, although one should
carefully study what would be the implications of one multiply connected metric for the other metric.

Moreover, since we are in a scenario where two different spaces are coexisting and
interacting only through gravitational effects, a bi-universe, one could wonder whether the impossibility of the simultaneous fulfillment of
the NEC for both effective stress-energy tensors in non-trivial cases is suggesting something deeper. In particular, it might be interesting
to consider that the physical matter coupled to one gravitational sectors fulfills the energy conditions, whereas 
the physical matter coupled to the other sector might
violate them (or fulfill their antithesis). In this case, it could be interesting to study if it would be possible
to formulate some kind of generalized quantum inequalities~\cite{Ford:1994bj} in the second space,  or even to consider 
whether the cosmic interest conjecture~\cite{Ford:1999qv} may be reversed in one space with respect to the other, with the ``quantum altruism'' conjecture being satisfied in the second universe~\cite{Pedro}.

\acknowledgments

VB acknowledges support by a Victoria University PhD scholarship.
PMM acknowledges financial support from the Spanish Ministry of
Education through a FECYT grant, via
the postdoctoral mobility contract EX2010-0854.
MV acknowledges support via the Marsden Fund and via a James Cook Research Fellowship, 
both administered by the Royal Society of New Zealand.

\appendix
\section{Some mathematical results regarding matrix square roots}\label{squareroots}
The matrix square root is defined through $\sqrt{A} \sqrt{A}=A$. Therefore, they fulfill
\begin{equation}
 (\sqrt{A})^{-1} = \sqrt{A^{-1}}; \qquad (\sqrt{A})^T = \sqrt{A^T}.
\end{equation}
Noting that
\begin{equation}
\sqrt{AB} B^{-1} = \sqrt{AB} B^{-1}  A^{-1} A = \sqrt{AB} (\sqrt{B^{-1} A^{-1}})^2 A = (\sqrt{AB})^{-1}  A,
\end{equation}
one can write
\begin{equation}\label{rel1}
\sqrt{AB} B^{-1} = (\sqrt{AB})^{-1}  A = \sqrt{B^{-1} A^{-1}}\; A.
\end{equation}
Following a similar procedure, we also have
\begin{equation}\label{rel2}
 A^{-1} \sqrt{AB} = B (\sqrt{AB})^{-1}  = B \sqrt{ B^{-1} A^{-1}}.
\end{equation}
Now, combining Equations (\ref{rel1}) and (\ref{rel2}), one has
\begin{equation}
 A^{-1} \sqrt{AB}  A = B \sqrt{AB} B^{-1}.
\end{equation}
This result can be made even stronger by noting that
\begin{equation}
( A^{-1} \sqrt{AB}  A)^2 =  A^{-1}( \sqrt{AB})^2 A = A^{-1} AB  A = BA, 
\end{equation}
and
\begin{equation}
( B \sqrt{AB}  B^{-1})^2 =  B ( \sqrt{AB})^2 B^{-1} = B AB   B^{-1} = BA.
\end{equation}
This leads to
\begin{equation}\label{rel3}
 A^{-1} \sqrt{AB}  A = \sqrt{BA} = B \sqrt{AB} B^{-1}.
\end{equation}
From (\ref{rel3})
\begin{equation}
 A^{-1} \sqrt{AB}  = \sqrt{BA}  A^{-1}, \qquad \hbox{and} \qquad  B \sqrt{AB}  = \sqrt{BA}  B.
\end{equation}
Re-naming $A\rightarrow A^{-1}$
\begin{equation}\label{e:key}
 A \sqrt{A^{-1}B}  = \sqrt{BA^{-1}}  A, \qquad \hbox{and} \qquad  B \sqrt{A^{-1}B}  = \sqrt{BA^{-1}}  B.
\end{equation}
which leads to
\begin{equation}
 (A \sqrt{A^{-1}B})^T  = A^T \sqrt{(A^{-1})^T B^T}, \qquad \hbox{and} \qquad  (B \sqrt{A^{-1}B})^T  =  B^T \sqrt{(A^{-1})^T B^T}.
\end{equation}
These are purely mathematical results holding for arbitrary not necessarily symmetric matrices $A$ and $B$. 

Relabeling $A\to g$ and $B\to f$, by iterating equations (\ref{e:key}) one can obtain
\begin{equation}
 g (\sqrt{g^{-1}f})^n = (\sqrt{fg^{-1}})^n  g,  \qquad \hbox{and} \qquad  f (\sqrt{g^{-1}f})^n = (\sqrt{fg^{-1}})^n  f.
\end{equation}
If we now take $f$ and $g$ to be symmetric tensors, then we have
\begin{equation}
( g (\sqrt{g^{-1}f})^n )^T= ((\sqrt{fg^{-1}})^n  g)^T = g (\sqrt{g^{-1}f})^n,
\end{equation}
and
\begin{equation}
( f (\sqrt{g^{-1}f})^n )^T= ((\sqrt{fg^{-1}})^n  f)^T = f (\sqrt{g^{-1}f})^n.
\end{equation}
The symmetry of these terms is what we have used to proof that both effective stress-energy  tensors are automatically symmetric.

As a side effect we also see
\begin{equation}
\gamma^T \; g \; \gamma = \left(\sqrt{g^{-1} f}\right)^T \; \left( g \; \sqrt{g^{-1} f}\right)  = \sqrt{f g^{-1}} \left( \sqrt{f g^{-1}} g \right) = f.
\end{equation}
That is
\begin{eqnarray}
f = \gamma^T \; g \; \gamma; \qquad \hbox{and} \qquad g = (\gamma^{-1})^T \; f \; \gamma^{-1}.
\end{eqnarray}
Furthermore, defining
\begin{equation}
S_{\mu\nu} =   g_{\mu\sigma} \; \gamma^\sigma{}_\nu, 
\end{equation}
which by the above is manifestly symmetric, we see
\begin{equation}
f_{\mu\nu} =  S_{\mu\sigma} \; g^{\sigma\rho} \; S_{\rho\nu},  
\qquad \hbox{that is} \qquad
f = S \; g^{-1} \; S,
\qquad \hbox{and} \qquad 
g = S \; f^{-1} \; S.
\end{equation}
This observation makes the $f\leftrightarrow g$ interchange symmetry between foreground and background very clear and explicit.

\section{Derivatives of the elementary symmetric polynomials}\label{theorem}
Let us consider the symmetric polynomials appearing in the interaction Lagrangian appropriate to 3+1 dimensions:
\begin{equation}
 L_\mathrm{int}=\alpha_1\,e_1(\gamma)+\alpha_2\,e_2(\gamma)+\alpha_3\,e_3(\gamma).
\end{equation}
These are
\begin{eqnarray}
e_1(X) &=&\tr[X];\\
e_2(X) &=&\frac{1}{2}\left(\tr[X]^2-\tr[X^2]\right);\\
e_3(X) &=&\frac{1}{6}\left(\tr[X]^3-3\tr[X]\tr[X^2]+2\tr[X^3]\right).
\end{eqnarray}
It can be seen that
\begin{equation}\label{deritrazas}
 \frac{\partial \tr[\gamma^n]}{\partial \gamma^\nu{}_\mu}=n\, \{\gamma^{n-1}\}^\mu{}_\nu ,
\end{equation}
with $\{\gamma^{0}\}^\mu{}_\nu =\delta^\mu{}_\nu $, $\{\gamma^{1}\}^\mu{}_\nu =\gamma^\mu{}_\nu $, $\{\gamma^{2}\}^\mu{}_\nu =\gamma^\mu{}_\sigma\gamma^\sigma{}_\nu $, and so on.
Therefore, we have
\begin{eqnarray}\label{derivadas}
 \frac{\partial e_1(\gamma)}{\partial \gamma^\nu{}_\mu}&=&\delta^\mu{}_\nu ;\\
 \frac{\partial e_2(\gamma)}{\partial \gamma^\nu{}_\mu}&=&\tr[\gamma]\delta^\mu{}_\nu -\gamma^\mu{}_\nu ;\\
 \frac{\partial e_3(\gamma)}{\partial \gamma^\nu{}_\mu}&=&
 \frac{1}{2}\left(\tr[\gamma]^2-\tr[\gamma^2]\right)\delta^\mu{}_\nu -\tr[\gamma]\gamma^\mu{}_\nu +\gamma^\mu{}_\sigma\gamma^\sigma{}_\nu .
 \label{e:deriv3}
\end{eqnarray}
Note that Equations (\ref{derivadas})--(\ref{e:deriv3}) can be written in a compact way as
\begin{equation}\label{deriv}
 \frac{\partial e_i(\gamma)}{\partial \gamma^\nu{}_\mu}=\sum_{m=1}^i (-1)^{m-1} \;e_{i-m}(\gamma)\;\{\gamma^{m-1}\}^\mu{}_\nu ,
\end{equation}
although (at this stage of the argument) this expression would only be justified for $i=\{1,\,2,\,3 \}$. We now prove that
equation (\ref{deriv}) holds for \emph{arbitrary} $n$. This is a necessary technical step in ultimately extending our argument to a 
$n$-dimensional Kaluza--Klein context.

Let our inductive hypothesis be that $\forall j\in\{1,2,3,\dots, k\}$ we assume
\begin{equation}\label{inductive}
 {\partial e_{i-j}(X)\over \partial X} = \sum_{m=1}^{i-j} \; (-1)^{m-1} \; e_{i-j-m}(X) \; X^{m-1},
\end{equation}
where we have omitted the indices for simplicity.
Differentiating the Newton identity
\begin{equation}\label{Newton}
 i \; e_i(X) = \sum_{j=1}^i (-1)^{j-1}  \; e_{i-j}(X) \; \tr[X^j],
\end{equation}
one obtains
\begin{equation}
i \; {\partial e_i(X)\over\partial X} =  \sum_{j=1}^i (-1)^{j-1}  \left\{ {\partial e_{i-j}(X) \over\partial X} \;  \tr[X^j]  + j \; e_{i-j}(X) \; X^{j-1}\right\}.
\end{equation}
The consideration of the inductive hypothesis (\ref{inductive}) in this expression leads to
\begin{equation}
i \; {\partial e_i(X)\over\partial X} =
\sum_{j=1}^i  \sum_{m=1}^{i-j}  (-1)^{j-1} (-1)^{m-1} \; e_{i-j-m}(X) \; X^{m-1}    \;  \tr[X^j]  +  \sum_{j=1}^i (-1)^{j-1}  j \; e_{i-j}(X) \; X^{j-1}.
\end{equation}
Note that the elementary symmetric polynomials are defined for positive subscript.
So, defining $e_{-1}=e_{-2}= e_{-3} \cdots = 0$, we can write
\begin{equation}
i \; {\partial e_i(X)\over\partial X} = 
\sum_{j=1}^i  \sum_{m=1}^{i}  (-1)^{j-1} (-1)^{m-1} \; e_{i-j-m}(X) \; X^{m-1}    \;  \tr[X^j]  +  \sum_{j=1}^i (-1)^{j-1}  j \; e_{i-j}(X) \; X^{j-1}.
\end{equation}
Taking the Newton identities (\ref{Newton}) into account, we can recognize the definition of $e_{i-m}(X)$ appearing in the first term
of the RHS. Thus, making this substitution and relabeling the sum of the second term, we have
\begin{equation}
i \; {\partial e_i(X)\over\partial X} =
\sum_{m=1}^{i} 
(-1)^{m-1} \;  (i-m) e_{i-m}(X) \; X^{m-1}   +  \sum_{m=1}^i (-1)^{m-1}  m \; e_{i-m}(X) \; X^{m-1},
\end{equation}
which clearly leads to
\begin{equation}
{\partial e_i(X)\over\partial X} = 
\sum_{m=1}^{i}  (-1)^{m-1} \; e_{i-m}(X) \; X^{m-1}.
\end{equation}
This proves the inductive step. Therefore, expression (\ref{deriv}) is true for arbitrary $n$.

\section{NEC in the $n$-dimensional theory}\label{ndimension}

We now consider the ghost-free interaction term for a $n$-dimensional bimetric gravity theory~\cite{Hinterbichler:2012cn}. 
Since $e_0(\gamma)=1$ and $e_n(\gamma)={\rm det}(\gamma)$, those two terms would correspond to a cosmological constant for $g$-space
and $f$-space, respectively. Therefore, we can absorb each cosmological constant in the kinetic term of the corresponding 
metric and write
\begin{equation}\label{intnsim}
 L_\mathrm{int}=\sum_{i=1}^{n-1} \alpha_i\,e_i(\gamma).
\end{equation}
Following a procedure similar to that in the 3+1 dimensional case, we write
\begin{equation}\label{taun}
 \tau^{\mu}{}_{\nu}=\sum_{i=1}^{n-1} \alpha_i\,\gamma^\mu{}_\sigma\,\frac{\partial e_i(\gamma)}{\partial \gamma^\nu{}_\sigma}.
\end{equation} 
As we prove in appendix~\ref{theorem}, the derivative terms fulfill 
\begin{equation}\label{derie}
 \frac{\partial e_i(\gamma)}{\partial \gamma^\nu{}_\mu}=\sum_{m=1}^i (-1)^{m-1} \; e_{i-m}(\gamma)\; \{\gamma^{m-1}\}^\mu{}_\nu .
\end{equation}
Thus we can re-write equation (\ref{taun}) as
\begin{eqnarray}\label{stressn}
 \tau^{\mu}{}_{\nu}=\sum_{i=1}^{n-1} 
 \sum_{m=1}^i \alpha_i (-1)^{m-1} e_{i-m}(\gamma) \, \{\gamma^{m}\}^\mu{}_\nu  .
\end{eqnarray}
The generalization of the NEC for $n$-dimensional $g$-space can now be expressed as
\begin{equation}\label{NECng1}
 \sum_{i=1}^{n-1}\sum_{m=1}^i \alpha_i\,(-1)^{m-1} e_{i-m}(\gamma)\,  \{ k_\mu \{\gamma^{m}\}^\mu{}_\nu  k^\nu\} \geq0,
\end{equation}
where $k$ is a null vector with respect to $g$.
Taking a null vector with respect to $f$, $\widebar{k}$, the NEC in $f$-space can be written as
\begin{equation}\label{NECnf1}
 \sum_{i=1}^{n-1}\sum_{m=1}^i \alpha_i\,(-1)^{m-1} e_{i-m}(\gamma)\, \{ \widebar{k}_\mu \{\gamma^{m}\}^\mu{}_\nu  \widebar{k}^\nu \} \leq0.
\end{equation}
Therefore, the NECs can be expressed in a very simple form even when considering a $n$-dimensional theory.
Now, noting that equation (\ref{violation}), that is
\begin{equation}
\bar k_\mu \,\{\gamma^m\}^\mu{}_\nu \, \bar k^\nu = k_\mu \, \{\gamma^m\}^\mu{}_\nu \,k^\nu,
\end{equation}
has been obtained without specifying the dimension of the matrices  involved, one can arrive to the same conclusion regarding
the simultaneous fulfillment of the NEC in both spaces. Therefore, the NEC is fulfilled in one and only one gravitational sector, if
it is not saturated.


\end{document}